\documentclass[reprint,amsmath,amssymb,apsm,siunitx,twocolumn]{revtex4-2}
\usepackage[utf8]{inputenc}
\usepackage{natbib}
\usepackage{color}
\usepackage{wasysym}
\usepackage{subfloat}
\usepackage{graphicx}
\usepackage{adjustbox} 
\usepackage{dcolumn}
\usepackage{breqn}
\usepackage{physics}
\usepackage{float}
\usepackage{algorithmic}
\usepackage[toc,page]{appendix}
\usepackage{tabularx}
\usepackage{steinmetz}
\usepackage{makecell, booktabs, multirow}

\newcommand{\otoprule}{\midrule[\heavyrulewidth]}
\usepackage{lipsum, fancyvrb}
\usepackage[caption=false]{subfig} 
\makeatletter
\renewcommand*{\fnum@figure}{{\normalfont\bfseries \figurename~\thefigure}}
\renewcommand*{\@caption@fignum@sep}{\textbf{ : }}
\makeatother

\begin{document}

\title{\textcolor{black}{Modeling of the driver transverse profile for laser wakefield electron acceleration at APOLLON Research Facility}}

\author{I. Moulanier$^1$$,$ L.T. Dickson$^1$$,$ C. Ballage$^1$$,$ O. Vasilovici$^1$$,$ A. Gremaud$^1$$,$ S. Dobosz Dufr$\text{é}$noy$^2$$,$ N. Delerue$^3$$,$ L. Bernardi$^4$$,$ A. Mahjoub$^4$$,$ A. Cauchois$^4$$,$ A. Specka$^4$$,$  F. Massimo$^1$$,$ G. Maynard$^1$$,$  B. Cros$^1$}
\affiliation{1 LPGP$,$ CNRS$,$ Universit$\text{é}$ Paris Saclay$,$ 91405 Orsay$,$ France}
\affiliation{2 CEA$,$ CNRS$,$ LIDYL$,$ Universit$\text{é}$ Paris Saclay$,$ 91190 Gif sur Yvette$,$ France}
\affiliation{3 IJCLAB$,$ CNRS$,$ Universit$\text{é}$ Paris Saclay$,$ 91405 Orsay$,$ France}
\affiliation{4 LLR$,$ CNRS/IN2P3$,$ Ecole Polytechnique$,$ Institut Polytechnique de Paris$,$ 91120 Palaiseau$,$ France}\rm{,}
\affiliation{*\rm{Corresponding authors : I. Moulanier}$,$ \rm{ioaquin.moulanier@universite-paris-saclay.fr}$;$ \rm{B. Cros}$,$ \rm{brigitte.cros@universite-paris-saclay.fr}}

\date{\today}

\begin{abstract}
The quality of electron bunches accelerated by laser wakefields  is highly dependant on the temporal and spatial features of the laser driver. Analysis of experiments performed at APOLLON PW-class laser facility shows that spatial instabilities of the focal spot, such as shot-to-shot pointing fluctuations or asymmetry of the transverse fluence, lead to charge and energy degradation of the accelerated electron bunch. 
It is shown that  PIC simulations can reproduce experimental results with a significantly higher accuracy when the measured laser asymmetries are included in the simulated laser's transverse profile, compared to simulations with ideal, symmetric laser profile. 
A method based on a modified Gerchberg-Saxton iterative algorithm is used to retrieve the laser electric field from fluence measurements in vacuum in the focal volume, and accurately reproduce experimental results using PIC simulations, leading to simulated electron spectra in close agreement with experimental results, for the accelerated charge, energy distribution and pointing of the electron beam at the exit of the plasma. 
\end{abstract}

\maketitle

\section{Introduction}

\vspace{0.5cm}

In the process of Laser WakeField Acceleration (LWFA)~\cite{Tagima1979,esarey2009physics}, an ultra-high intensity laser is focused inside a gas target, ionizes the medium and creates a trailing perturbation in its wake in an underdense plasma. 
The generated plasma cavity sustains intense longitudinal and transverse electric fields that can trap, accelerate and focus bunches of electrons to the GeV range~\cite{tsung2004near} within a few centimeters.
Injection of plasma electrons makes LWFA a compact option for the generation of relativistic electron sources. However, despite numerous studies, the use of electron beams from LWFA for applications is impeded by insufficient beam quality and stability. Therefore, \textcolor{black}{detailed diagnostics}, realistic modeling and analysis are needed to achieve a precise understanding of the key mechanisms controlling the laser plasma interaction in experiments.

The main schemes for the injection of plasma electrons into the plasma wave are self-injection~\cite{corde2013observation} and ionization injection~\cite{chen2012theory, mcguffey2010ionization, pak2010injection, mirzaie2015demonstration}. \textcolor{black}{In LWFA experiments with PW-class laser drivers}, both injection schemes \textcolor{black}{can} occur in the so-called bubble regime~\cite{pukhovBubble2002}, in which the ponderomotive force of the laser repels plasma electrons from its propagation axis, generating an electron-free cavity behind the laser pulse. In the process of self-injection, a portion of the expelled plasma electrons travel around the cavity before getting trapped in the wakefield~\cite{kostyukovTransverInjection2009}. In ionization injection, the gas target is a mixture of light atomic species, typically hydrogen, ionized early before the peak of the laser pulse, and a dopant species, e.g. nitrogen, presenting an energy-gap in its ionization potential structure~\cite{pak2010injection}, leading to ionization of some electrons close to the peak of the laser pulse.

With peak intensities above $I_0=10^{18}$ W/cm$^{2}$, the pulse temporal front ionizes  hydrogen and nitrogen up to $N^{5+}$. Remaining nitrogen L-shell electrons  are primarily born  around laser peak intensities, inside the bubble~\cite{massimo2020numerical}. 
Ionization injection has several properties of interest for tuning electron injection and trapping, and favors highly charged electron beams. It operates at an intensity below self-injection~\cite{mirzaie2015demonstration} and the two mechanisms can be optimized in different parameter areas. 
As ionization injection depends on the local intensity of the laser pulse, it can be particularly sensitive to laser beam quality and its evolution during propagation in the evolving plasma density. These properties can be used to control the injection process in electron density tailored profile  or diagnose laser beam quality.

The electron beam charge can be increased by increasing the driving laser power, providing a large range of parameters to explore for optimizing the properties of laser driven electron sources with PW class short pulse laser facilities.

This complex nonlinear physics is  described using particle in cell (PIC) simulations \cite{BirdsallLangdon2004}, using as input parameters the laser temporal and spatial shapes, and the gas density profile. Experimentally achieved laser beams often differ from perfectly symmetrical distributions. Hence, in order to understand the role of  laser imperfections on the quality of the produced electron bunches, and compare to experimental results, refined PIC simulations describing realistically the injection and acceleration physics occurring at the ps scale were performed. 
These realistic simulations require a proper description of the gas density  profile, as well as of the driver (laser profile transverse asymmetries) to accurately reproduce  laser-plasma interactions affecting the electron beams  characteristics.
For instance, laser asymmetries have been shown to lead to asymmetric wakefields affecting the output accelerated electron beam, with characteristics  directly correlated with the laser stability and quality\textcolor{black}{~\cite{Beaurepaire2015,ferri2016effect, Dickson2022, leemans2014multi, popp2010all}}.

In this paper the method used to analyze characteristic results obtained during commissioning experiments at APOLLON PW facility~\cite{papadopoulos2016apollon} is presented. Focusing the $\mathrm{F_2}$ laser  beam at 0.4~PW in the long focal area  inside a  gas cell~\cite{audet2018gas},   experiments were performed to
characterize  laser beam quality and evaluate its impact on electron properties. 

It is shown that PIC simulations can reproduce experimental results with a significantly higher accuracy when the measured laser asymmetries are included in the simulated laser's transverse profile. This enhanced agreement is meant in comparison with simulations with an ideal, axisymmetric laser profile, which is often used in the design stage of LWFA experiments and in preliminary experimental analyses. The results described in this work thus show the importance of more realistic initial conditions in numerical modeling used for these studies.
The simulation results shown in this work were obtained with the quasi-3D PIC code FBPIC~\cite{lehe2016FBPIC}, but the same method can be applied with other PIC codes in quasi-3D \cite{LIFSCHITZ20091803} or full 3D geometry.

In comparison to previous investigations made with realistic PIC simulations in~\cite{ferri2016effect}, this work uses an alternative fast Gerchberg-Saxton algorithm~\cite{gerchberg1972practical} \textcolor{black}{based on mode decomposition} to reconstruct the laser field, which allows the accurate simulation of an experimental electron bunch spectrum in the energy-angle plane. This method has already been used to present experimental results in~\cite{Dickson2022}, \textcolor{black}{in a regime with a lower peak laser intensity and characterized by a more stable transverse laser profile from shot-to-shot}. In addition, in this work, the physical effects of the realistic laser driver (in particular its asymmetries) on the electron injection in the bubble are described. An agreement between realistic numerical modeling and experiment is obtained also in the electron beam spectra in the energy-angle plane. Furthermore, the realistic simulations in \cite{ferri2016effect} have been performed in 3D, while the realistic simulations of this work were performed in quasi-3D geometry \cite{LIFSCHITZ20091803}, highlighting that high-fidelity simulations can be obtained also with this less computationally-demanding technique for preliminary analyses.

The remainder of this paper is organized as follows.
Characteristic experimental results  are presented in section II. The method used to retrieve the \textcolor{black}{laser} electric field from experimentally recorded fluence images is described in section III, followed by the description of the method to generate data to initialize FBPIC simulations. A comparison of experimental and numerical electron spectra is discussed in section IV.

\section{Experimental results}

An experiment was performed in April 2021 during the commissioning phase of the long focal area of APOLLON facility to characterize laser beam quality inside the experimental area and evaluate its impact on electron beam quality.

\begin{figure}[h!]
     \begin{center}
          \includegraphics[width=9cm]{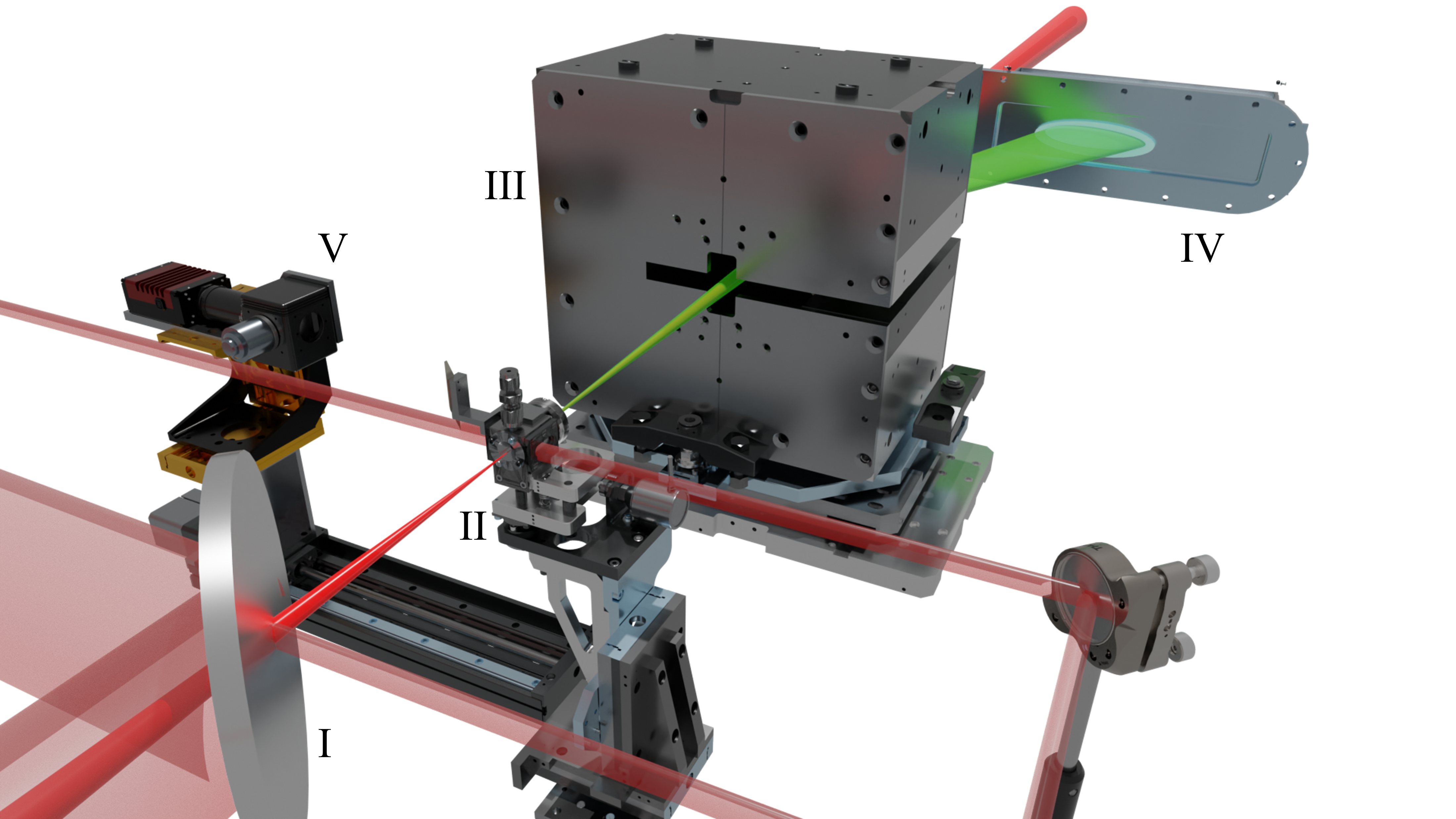}
     \end{center}
     \caption{Schematic of experimental set-up : the driver laser beam (in red) is focused through the turning mirror (I) at the entrance of the gas cell (II), generating a diverging electron bunch (in green) at the exit. The electron bunch is then deflected by a magnetic dipole (III) and sent onto a LANEX screen (IV) to measure its Energy-Angle distribution.
     A small percentage of the \textcolor{black}{laser} driver is used to probe the plasma density transversely.
     Vacuum imaging of the focal volume is completed with a CCD (V).
     }
     \label{fig:setup}
 \end{figure}

After compression, the $\mathrm{F_2}$ \textcolor{black}{laser} beam was transported into the experimental area and focused in vacuum using an on-axis spherical mirror $3$ m focal length, after reflection from a turning mirror with a hole, as illustrated in Fig.~\ref{fig:setup}.

The central part of the \textcolor{black}{laser} beam  incident on the turning mirror was collected and used partly to monitor the laser beam energy from shot-to-shot  using a leak through a wedge and a calorimeter. The remainder of the \textcolor{black}{laser} beam was used as a probe \textcolor{black}{laser} to diagnose plasma density transversely to the main pulse. The relative probing time was controlled by an in vacuum delay stage.

For this experiment, the APOLLON F$_2$ $\mathrm{Ti:Sa}$ laser with a central wavelength $\lambda=0.8$, $\mu$m was measured to deliver a pulse with a FWHM intensity duration $\tau_{FWHM} = 25$ fs, a post-compression energy $E_l=5$ to $10$ J, with a repetition rate of 1 shot per minute.
The peak intensity estimated in the ideal Gaussian transverse profile approximation in vacuum is $5 \times 10^{19}$ W/cm$^{2}$. The \textcolor{black}{laser} beam was focused  inside a $6$ mm long gas cell~\cite{audet2018gas} filled with a mixture of $99\%\mathrm~{H_2}$ and $1\%\mathrm~{N_2}$. Electrons were trapped  through ionization and accelerated in the wakefield. After plasma exit, their energy was measured using a dipole magnet and $\mathrm{LANEX}$ screen imaged onto a $\mathrm{CCD}$ camera. The spectrum was recorded in the $300-900$ MeV energy range within a $\pm 20$ mrad viewing angle.

\begin{figure}

    \includegraphics[width = \linewidth]{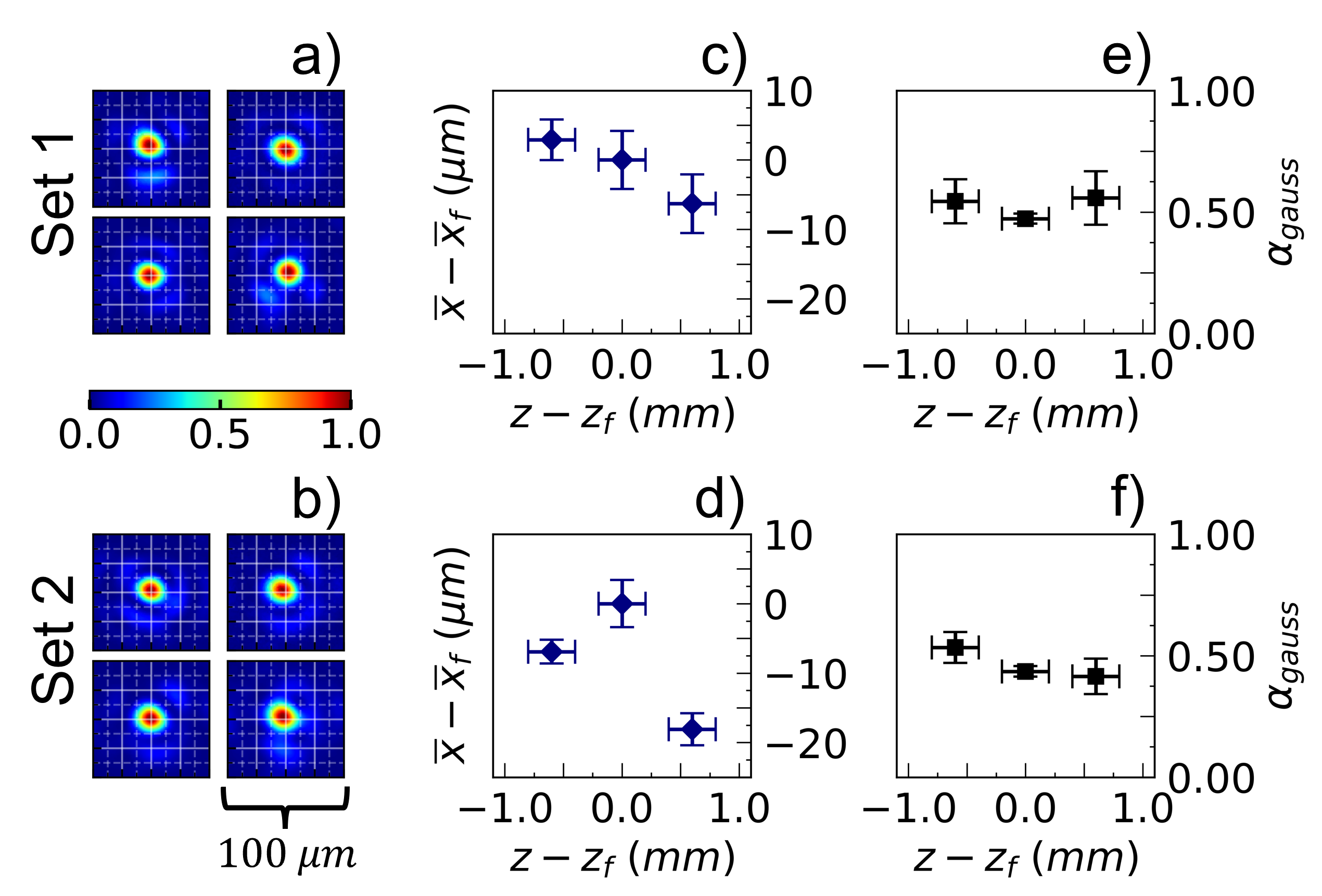}
    \caption{Measured laser beam stability during 2 sequences of multiple vacuum shots taken on 2 separate days with  laser settings 1 and 2:
    (a) and (b) laser fluence of 4 consecutive shots measured in the focal plane;
    (c) and (d) $\overline{x} - \overline{x}_f$, \textcolor{black}{laser} beam relative centroid in the $xOz$ plane, with respect to the focal plane centroid, for three positions $z - z_f$ along the \textcolor{black}{laser} propagation axis; the data for $z=z_f$ is from the 4 fluence images of (a) and (b), while the other points are data collected from 2 separate shots at each position; 
    (e) and (f) fraction of  energy $\alpha_{gauss}$ inside a Gaussian fit of waist $w_0$. \textcolor{black}{Error bars along $z$ come from the determination of the focal plane $z_f$.} 
   }
    \label{fig:laser_exp}

\end{figure}

The laser beam was characterized in detail in the focal volume every day prior to shots on the gas-filled target. 
Using a movable CCD camera in vacuum, the  laser energy distribution was measured  inside the interaction chamber before, at and after focus. The Rayleigh length in vacuum, defined as $z_R = (\pi w_0^2) / \lambda$, with $w_0$ being the $1/e^2$ radius of the focal plane intensity, was $z_R=1$ mm. The waist of a Gaussian best fit in the focal plane is $w_0=16.6\pm0.3$ $\mu$m. 
Fig. \ref{fig:laser_exp} shows  2 sets of data taken on two different days, illustrating instabilities both in laser pointing and in spatial fluence symmetry between consecutive shots and for different days. 
Fig. \ref{fig:laser_exp}.(a) and (b) show  the fluence distribution in the focal plane $z = z_f$.

Fig.~\ref{fig:laser_exp}.(c) and (d) show the \textcolor{black}{shot-to-shot} transverse centroid displacement fluctuations $\overline{x} - \overline{x}_f $ in the focal volume \textcolor{black}{(vertical error bars)}. \textcolor{black}{The horizontal error bars come from the determination of the focal plane $z_f$.}
The $xOz$ plane is the plane where electrons centroid fluctuations have been measured, perpendicularly to the laser polarization plane $yOz$.
Fig.~\ref{fig:laser_exp}.(c) and (d) underline instabilities of the laser centroid in both cases. For Set 1 data   the centroid is moving linearly, going from $+3$ $\mu$m at $z-z_f=-0.6$ mm to $-6$ $\mu$m at $z-z_f=+0.6$ mm.
Set 2 data show larger fluctuations $\overline{x} - \overline{x}_f =-7$ $\mu$m at $z-z_f=-0.6$ mm and $\overline{x} - \overline{x}_f =-18$ $\mu$m at $z-z_f=+0.6$ mm. Multi-directional fluctuations of the centroid around the focal plane are the signature of a non zero temporal phase and the asymmetry of the laser fluence.

In Fig.~\ref{fig:laser_exp}.(e) and (f), $\alpha_{gauss}$, defined as the fraction of total  energy inside a Gaussian fit with  waist $w_0$,  is plotted as a function of position in the focal volume.  Values of $\alpha_{gauss}$ at $z \neq z_f$ are calculated using a waist $w(z)=w_0 [1+\left(z/z_R\right)^2]^{1/2}$ for the Gaussian fit. At $z=z_f$, $\alpha_{gauss}$ is averaging $46\,\%$ for Set 1 and $40\,\%$ for Set 2. This demonstrates that the experimental fluence departs significantly from a perfect Gaussian approximation in both cases.

\begin{figure}[ht]
        \includegraphics[width = \linewidth]{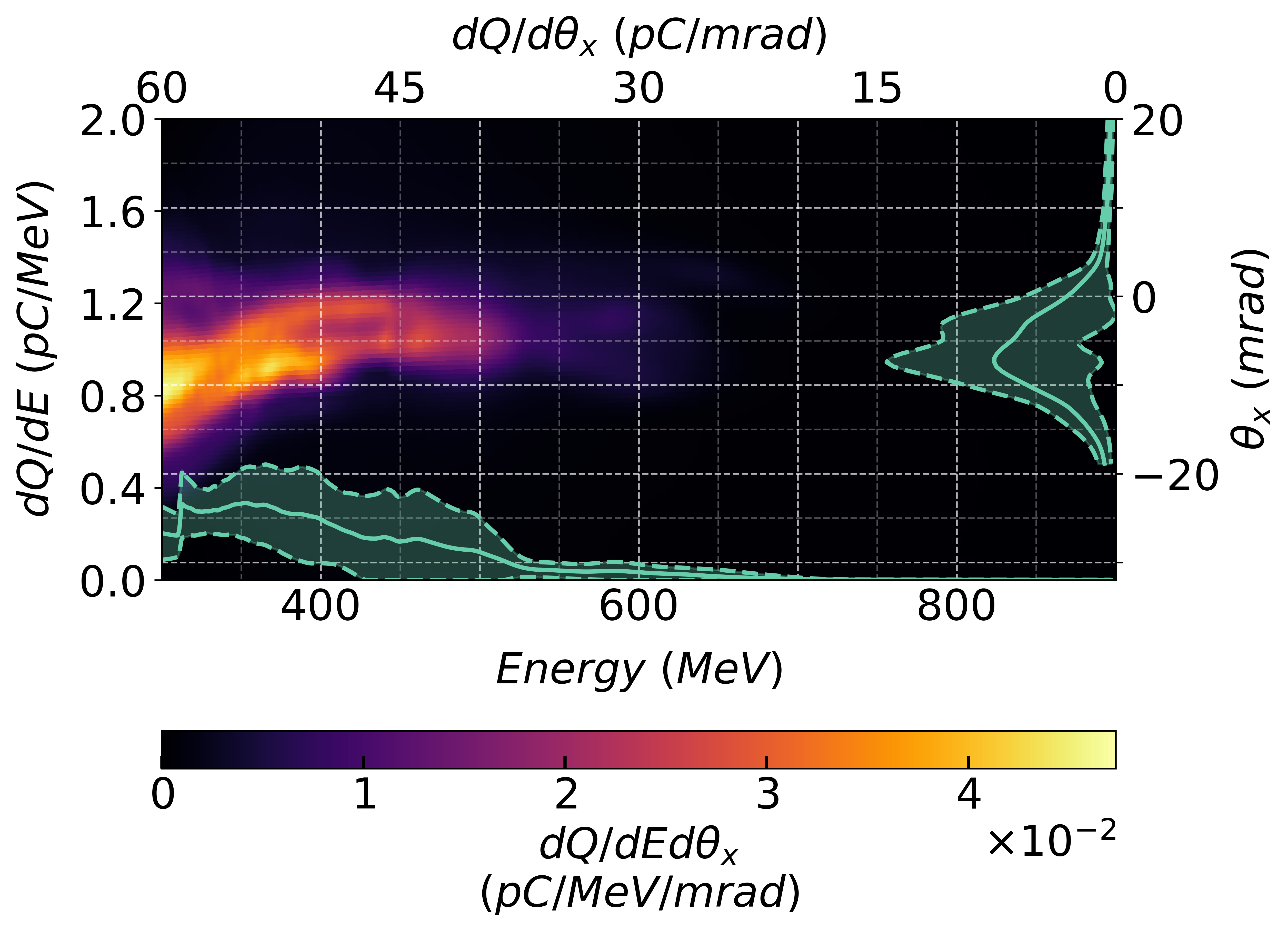}
    \caption{Average electron spectrum of 10 consecutive shots acquired the same day as Fig.~\ref{fig:laser_exp} Set 1 laser measurements for an average plasma electron density $\overline{n}_{0} = 2.2\times\mathrm{10^{18}}$ cm$^{-3}$. The color scale shows the spectral charge density in the $Energy-\theta_x$ angle plane. The full green line along the $Energy$ axis is the average spectral charge density, $dQ/dE$, profile (scale on left vertical axis) calculated within $\pm 3$ mrad around each individual maximum $dQ/d\theta_x$. The full green line along the $\theta_x$ axis (right-handside vertical axis) represents the average $dQ/d\theta_x$ profile (scale on the top horizontal axis) integrated over energy. The green areas are the confidence intervals bounded by the standard deviation extremes (dashed curves). $\theta_x = 0$ corresponds to the laser axis alignment position in vacuum.}
    \label{fig:spectra_exp}
\end{figure}

Figure~\ref{fig:spectra_exp} shows the average of 10 consecutive electron bunch spectra  measured on the same day as   Fig.~\ref{fig:laser_exp} Set 1, in $yOx$, also defined as the $Energy-\theta_x$ angle plane.
The average bunch charge measured between $300$ and $900$ MeV and within $\pm 20$ mrad is $\overline{Q}_{tot} = 95 \pm 46$ pC. For this sequence, the mean measured laser energy is $\overline{E}_{l} = 4.8\pm0.2$ J, and average plasma electron density $\overline{n}_{0} = 2.2\pm0.2\times\mathrm{10^{18}}$ cm$^{-3}$. The measured charge fluctuations are $\delta\overline{Q}_{tot}/\overline{Q}_{tot}=47\,\%$, which shows a clear sensitivity to input parameters.

\section{Modeling of the laser beam}
The analysis and implementation of the laser experimental data into PIC simulations are performed in three steps. First, the \textcolor{black}{laser} electric field is retrieved from the measured fluence data. Then, the reconstructed \textcolor{black}{laser} electric field is represented as a mode sum of fields that are used as parameters at the start of PIC simulations. Finally, the calculated electron parameters at the end of each simulation are compared with measured results.

\subsection{Fit of the \textcolor{black}{laser} electric field}
A modified version of the reconstructive Gerchberg-Saxton algorithm (GSA) \cite{gerchberg1972practical}, described in Appendix \ref{appendix:GSA_fit}, was used to retrieve the \textcolor{black}{laser} electric field corresponding to the laser fluence measured at 3 positions : $z - z_f = 0,\,+1.2,\,-1.8$ mm. 

The algorithm was used in particular to retrieve an unknown phase map, $\psi(x,y)$, associated to a set of fluence images measured at different positions $\left( z_0,\,z_1...\,z_{k_{max}}\right)$.

\begin{figure}

    \includegraphics[width = 0.85\linewidth]{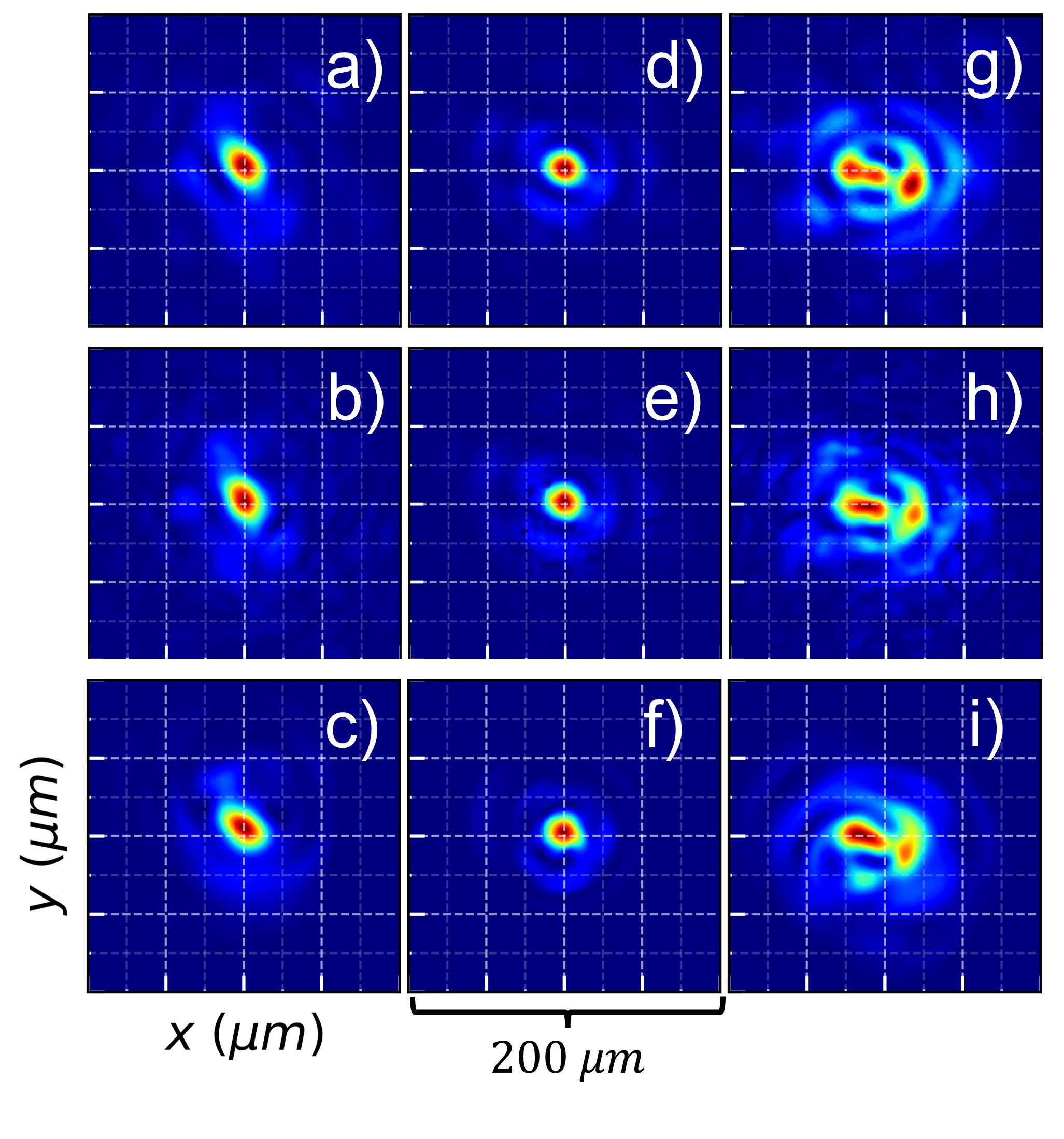}
    \caption{\textcolor{black}{Experimental laser fluence measured in the focal volume (upper row), corresponding Hermite-Gaussian reconstruction (middle row), Laguerre-Gauss fit with $L=3$ and $M=40$ used in simulations (lower row): 
    (a), (b) and (c) $z-z_f=-1.8$ mm 
    - (d), (e) and (f) $z-z_f=+0$ mm 
    - (g), (h) and (i) $z-z_f=+1.2$ mm.
    Each distribution has been normalized by its peak fluence.}}
    \label{fig:GSA_fit}
    
\end{figure}

We selected laser data from the same day as Fig.~\ref{fig:laser_exp} Set 1, which exhibit  better focal spot stability than Set 2, to reconstruct a realistic laser electric field distribution. For each position, the corresponding fluence distribution has been re-centered around the origin in order to reduce in advance the shot-to-shot fluctuations error on the fit.
The selected experimental distributions (upper row) and results of the fit algorithm (\textcolor{black}{middle row}) are plotted in Fig.~\ref{fig:GSA_fit}. 

Using a combination of low order fits to establish an educated guess, then refining the optimization with a higher order Hermite-Gaussian modes projection, a realistic reconstructed distribution is calculated as represented in Fig.\ref{fig:GSA_fit}, \textcolor{black}{(middle row)}. 

\textcolor{black}{The laser fluence reconstructed with Laguerre-Gauss modes is plotted in Fig.~\ref{fig:GSA_fit} (lower row) and shows a relatively good agreement between the reconstructed Laguerre-Gauss distribution and the experimental laser data.}

\subsection{PIC simulation set-up}\label{sec:pic_sim_setup}
Due to the cylindrical representation used in FBPIC, the reconstructed Hermite-Gauss \textcolor{black}{laser} electric field $\mathrm{E}_{\mathrm{HG}}(r,\theta,z_f)$ is projected on Laguerre-Gauss modes~\cite{pampaloni2004Gaussian} in the focal plane $z=z_f$ :
\begin{equation}
    \begin{split}
    C_{l,m} = \int_{r=0}^{r_{max}}&\int_{\theta=0}^{2\pi} \mathop{r} \mathop{dr} \mathop{d\theta} ~ \mathrm{E}_{\mathrm{HG}}(r,\theta,z_f) \\ &\times \mathrm{LG}_{l,m}^*(r,\theta,z_f)(r, \theta, z_f, r_{0,opt}, \theta_{0,opt}, w_{0r})
    \end{split}
\end{equation}
with $C_{l,m}$ being the complex amplitudes of the $\mathrm{LG}_{l,m}$ Laguerre-Gauss modes, $l$ the azimuthal order, $m$ the radial order, $^*$ the complex conjugate operator, $w_{0r}=(w_{0x}^2+w_{0y}^2)^{1/2}$ the projection waist, $\left(r_{0,opt}, \theta_{0,opt}\right)$ the GSA cycle optimized origins in cylindrical coordinates at $z=z_f$ and $r_{max}=\min\left( \Delta X/2, \Delta Y/2\right)$. The rectangular grid length along each axis is denoted by $\left( \Delta X,\Delta Y \right)$.

 For a given number $M$ of radial modes, the quality of the Laguerre-Gauss projection as a function of the number of azimuthal modes taken into account (azimuthal order $L$) is evaluated by calculating $\epsilon_{fit}$, the integral error, defined as :
\begin{equation}
    \epsilon_{fit} = \dfrac{\int\displaylimits_{r=0}^{r_{max}}\int\displaylimits_{\theta=0}^{2\pi} \mathop{r}\mathop{dr}\mathop{d\theta}\left|F_{\mathrm{HG}}(r,\theta, z_f)-F_{\mathrm{LG}}(r,\theta, z_f)\right|}
    {\int\displaylimits_{r=0}^{r_{max}}\int\displaylimits_{\theta=0}^{2\pi} \mathop{r}\mathop{dr}\mathop{d\theta} F_{\mathrm{HG}}(r,\theta, z_f)}
\end{equation}
with $F_{\mathrm{LG}}(r,\theta, z_f)$ being the normalized fluence of the Laguerre-Gauss modes, $F_{\mathrm{HG}}(r,\theta, z_f)$ the normalized fluence of the Hermite-Gauss modes from the \textcolor{black}{laser} electric field fit, both in the focal position $z=z_f$.

The integral error $\epsilon_{fit}$ decreases with the integration radius $r_{max}$ for a fixed set of ${\mathrm{LG}}$ modes. 
$\epsilon_{fit}$ was calculated for an effective interaction radius, $r_{eff}$, to evaluate the quality of  the Laguerre-Gauss fit near the peak fluence of the reconstructed profile. As  injection and acceleration of electrons both occur within a characteristic radius around the laser centroid, on a  scale of the order of the bubble radius $R_b\propto w_0$~\cite{lu2007generating},  we set $r_{eff}=R_b$ and evaluate the error for  $r\leq r_{eff}$ to measure the accuracy of the fit.
In PIC simulations performed with a perfect Gaussian laser, the plasma cavity has values of $R_b$ ranging from $15$ to $20$ $\mu$m. Therefore, we set $r_{eff}=20$ $\mu$m as the value corresponding to the maximum observed bubble radius.

\begin{figure}[t]

    \includegraphics[width=0.9\linewidth]{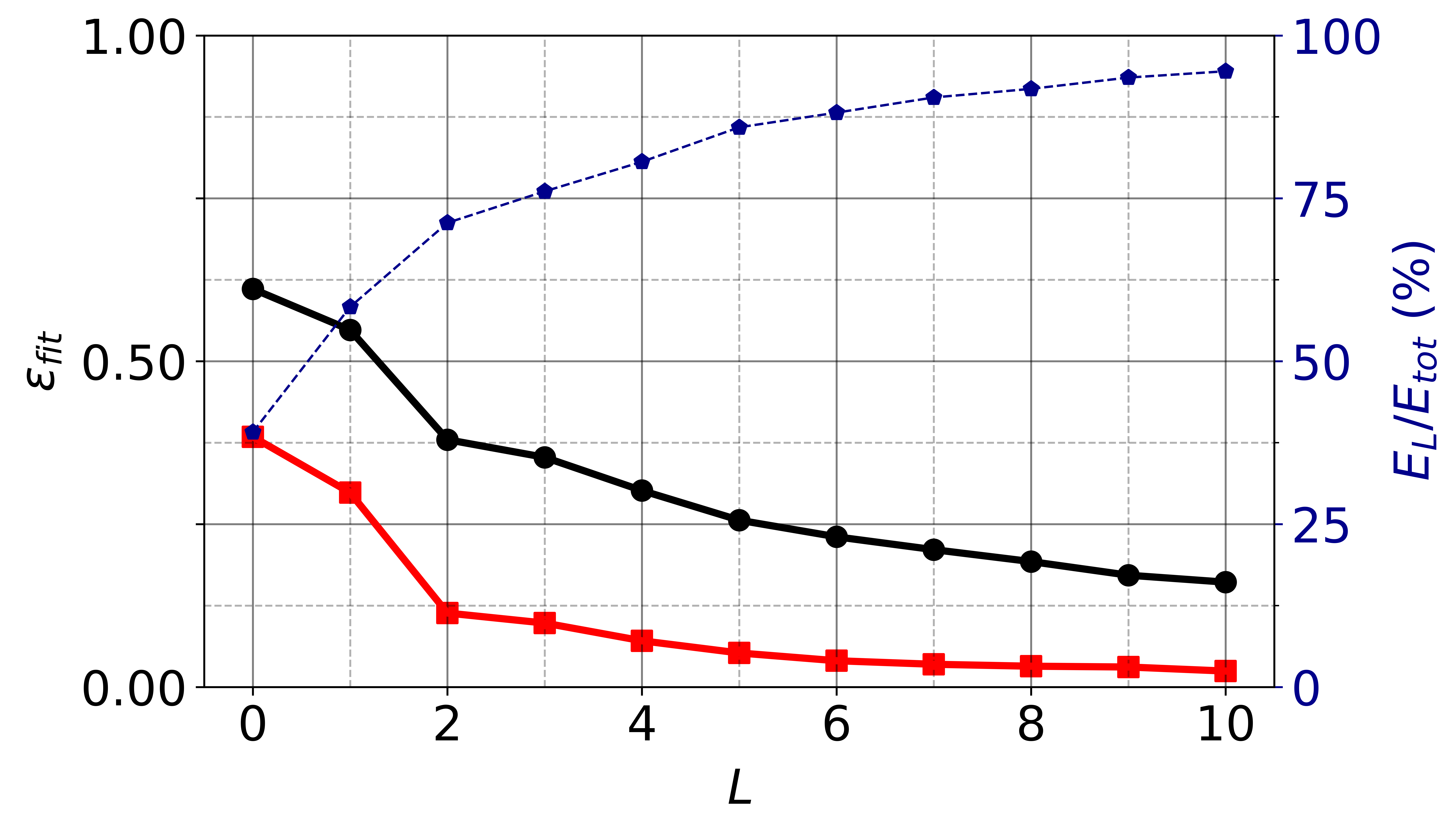}
    \caption{Integral error $\epsilon_{fit}$ for $M=40$ radial modes, as a function of the maximum number of azimuthal modes $L$. Red squares: $\epsilon_{fit}$ calculated in a disk of radius $r_{eff}=20$ $\mu$m; black circles: $\epsilon_{fit}$ calculated in a disk of radius $r_{max}=200$ $\mu$m; blue diamonds (vertical axis on the right): cumulative energy fraction included in  each configuration. }
    \label{fig:LG_fit}
    
\end{figure}

Figure~\ref{fig:LG_fit} shows the evolution of $\epsilon_{fit}$ for $M=40$ and $L$ ranging from $0$ to $10$. 
The error of the fit converges toward $\epsilon_{fit}=0$ both within $r_{eff}$ and $r_{max}=200$ $\mu$m as the number of azimuthal modes used for projection increases. A share of $43\,\%$ of the total energy is contained within the first mode $L=0$ within a radius $r_{eff}$, which shows that the remainder is contained within non-symmetrical azimuthal modes. 

The number of azimuthal modes for the Laguerre-Gauss harmonics needs $N=5$ azimuthal modes to be simulated with the azimuthal harmonics of type $\exp{-in\theta}$ used in quasi-3D geometry \cite{LIFSCHITZ20091803}. In addition, the number of macro-particles per cell in the simulations in this work has been increased with the number of azimuthal modes to maintain a constant signal to noise ratio (whose value in quasi-3D simulations is discussed in \cite{LIFSCHITZ20091803}). As a result, the computational time required for a simulation with a Laguerre-Gauss laser field is increased due to the increased number of modes and of macro-particles per cell.
\textcolor{black}{Every simulation in this work with a Laguerre-Gauss sum laser profile used $L=3$ and $M=40$, which reproduces 75\% of the total energy as shown in Fig.~\ref{fig:LG_fit}. For $L>$3, the fit error $\varepsilon_{fit}$ decreases slowly compared to the increase in computational time associated with the corresponding number of modes.}

All numerical results presented in this paper were obtained with input parameters in the same range as those of the data shown in Fig.~\ref{fig:spectra_exp}.
The simulated laser beam has a FWHM duration $\tau_{FWHM,sim}=25$ fs (Gaussian temporal profile) and the focal plane fluence is fitted with a Gaussian waist $w_{0,sim}=16$ $\mu$m.
The laser is propagated over $10$ mm in a gas cell filled with $99\,\%\mathrm{H_2}-1\,\%\mathrm{N_2}$ gas, including a starting $1.4$ mm up-ramp from $n = 0$ to a density plateau of $n = n_{0,sim}$, as well as a down-ramp to $n = 0$ from $ct = 7.4$ mm to $ct = 10$ mm. \textcolor{black}{The plasma density longitudinal profile profile is inferred from OpenFOAM simulations~\cite{audet2018gas}}. For both Gaussian and realistic simulations, the laser driver is focused onto the start of the density plateau at $z_f=1.4$ mm.

\textcolor{black}{For simulations using the Gaussian profile, the energy was fixed at $E_{l,sim}=\alpha_{gauss}\times E_l$, with $\alpha_{gauss}=0.46$ the value calculated in Fig. 2.(e) at $z=z_f$.
For simulations using the Laguerre-Gauss profile, simulated with $L=3$ and $M=40$, $E_{l,sim}=0.75 \times E_l$.}

The simulation grid is represented along $z$ by $N_z=3000$ points with an increment $\Delta z=0.025$ $\mu$m, and along $r$ by $N_r=1100$ points with an increment $\Delta r=0.2$ $\mu$m.

In the simulations with the  reconstructed laser field profile, each population of $\mathrm{H_2}$ and $\mathrm{N_2}$ was simulated with $[P_z,P_r, P_\theta]=[2,2,16]$ macro-particles per cell along $z$, $r$ and $\theta$ respectively. In the following, these simulations will be referred to as "realistic simulations".

Instead, every simulation using a perfect Gaussian laser profile was performed with $N_m=2$ azimuthal modes. For these simulations, each population of $\mathrm{H_2}$ and $\mathrm{N_2}$ was simulated with $[P_z,P_r, P_\theta]=[2,2,4]$ macro-particles per cell.

\section{Realistic PIC simulations results}

\subsection{Influence of laser asymmetry on electron beam spectra}\label{sec:realistic_sim}

To understand the physical impact of laser asymmetry on the electron bunch quality,  realistic simulations were performed with FBPIC using the reconstructed laser driver retrieved from  fluence distributions shown in Fig.~\ref{fig:GSA_fit}. 
Input parameters were set as described in the previous section, and \textcolor{black}{the electron density was fixed to $n_{0,sim} = 2.1\times\mathrm{10^{18}}$ cm$^{-3}$.} \\

\begin{figure}[ht!]
    \centering
     \includegraphics[width = 0.9\linewidth]{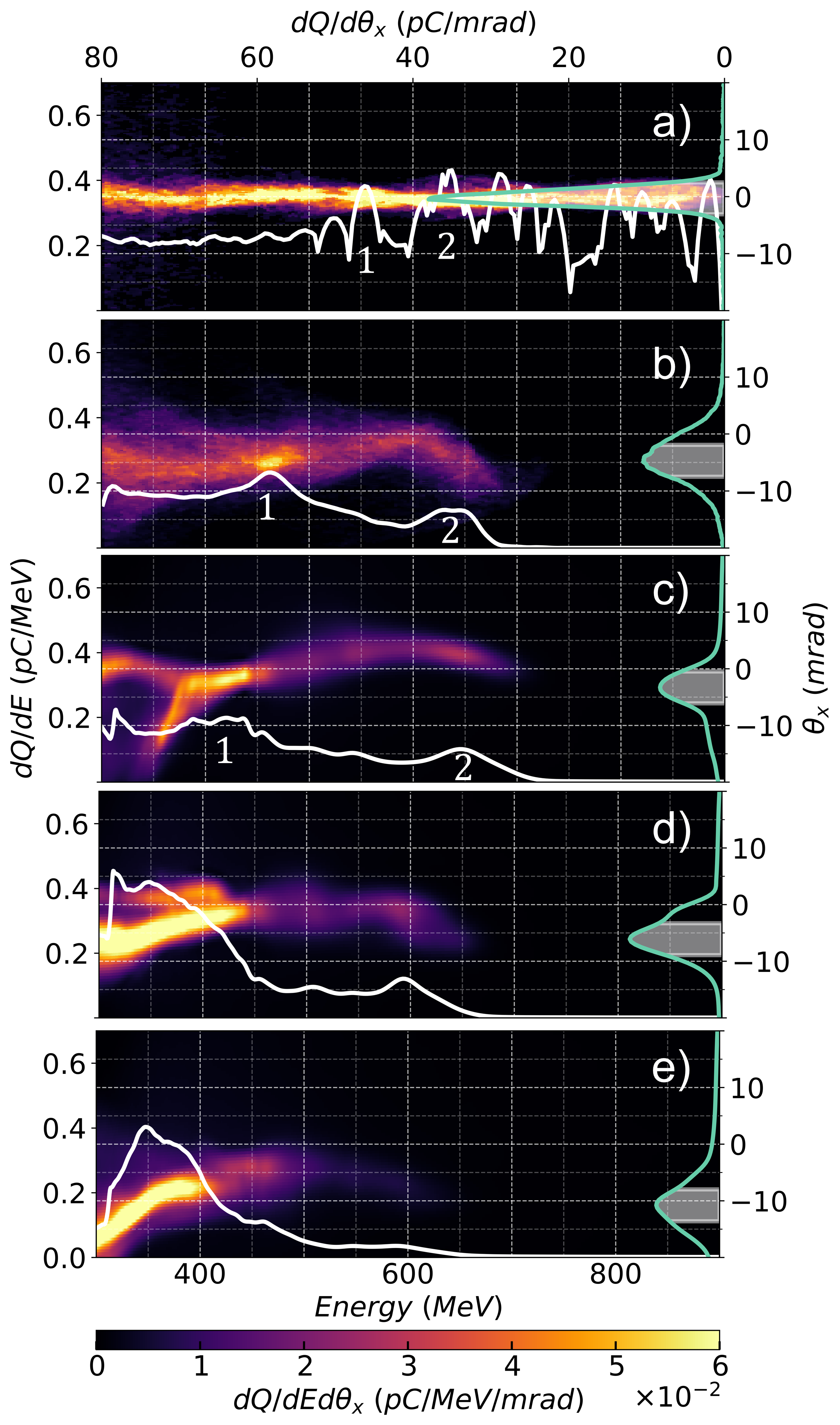}
    \caption{\textcolor{black}{Electron bunch $dQ/dEd\theta_x$ spectral density in the energy~$\theta_x$~divergence plane for 3 different cases : (a) simulated Gaussian spectrum - (b) simulated Laguerre-Gauss spectrum - (c) to (e) single experimental spectra from Fig.~\ref{fig:spectra_exp}.(a) sequence. White (green) lines represent spectral charge, $dQ/dE$, (divergence charge density $dQ/d\theta_x$) integrated over the divergence (energy) 
 respectively. The shaded area under the green $dQ/d\theta_x$ line shows the range $\pm 3$ mrad around the peak of $dQ/d\theta_x$, used to produce the $dQ/dE$ curve and calculate the spectrum peaks properties.
    The label 1 and 2 reference specific peaks within the $dQ/dE$ profiles corresponding to Table~\ref{fig:table_spectra_sim}.}}
    \label{fig:spectra_sim}
    
\end{figure}

For the sake of comparison between simulation and experimental results, 5 electron spectra are shown in Fig.~\ref{fig:spectra_sim}: (a) simulation with a Gaussian laser driver, (b) simulation with a Laguerre-Gauss laser driver  using distribution described by Figs.~\ref{fig:GSA_fit} and \ref{fig:LG_fit}, both with electron density $n_{0,sim} = 2.1\times\mathrm{10^{18}}$ cm$^{-3}$,  and (c) to (e) experimental spectra (single instances of the data shown in Fig.~\ref{fig:spectra_exp}) measured with electron densities \textcolor{black}{$n_{0} = 2.1\times\mathrm{10^{18}}$ cm$^{-3}$ for (c) and (e),  $n_{0} = 2\times\mathrm{10^{18}}$ cm$^{-3}$ for (d) and laser energy $E_l=4.7$ J for (c) to (e).
The total charge of the average of the experimental spectra from (c) to (e) is 111 pC, and their average central divergence is $-6.6$ mrad.}

\begin{table*}[ht!]

  \renewcommand{\arraystretch}{1.3}
  \begin{tabular}{c c c c c c c c}
    \toprule
\textbf{Spectrum} & $Q_{tot}$ & $\theta_{x, max}$ & \textbf{Peak Label} & $E_{peak}$ & $\Delta E_{peak}/E_{peak}$ & $Q_{peak}$ &  $dQ/dE\,max.$\\
    \otoprule
    \textbf{Units} & pC & mrad & & MeV & $\mathrm{\%}$ & pC &  pC/MeV\\
    \midrule
    \multirowcell{2}{\textbf{Simulation} \\ \textbf{Gauss}} & \multirowcell{2}{\textcolor{black}{174}} & \multirowcell{2}{-0.3} & \textbf{1} & 556 & \textcolor{black}{3.2} & \textcolor{black}{7} & \textcolor{black}{0.38} \\ & & & \textbf{2} & \textcolor{black}{629} & \textcolor{black}{3.3} & \textcolor{black}{8} & \textcolor{black}{0.41} \\
    
    \midrule
    \multirowcell{2}{\textbf{Simulation} \\ \textbf{Laguerre-Gauss}} & \multirowcell{2}{109} & \multirowcell{2}{-4.7} & \textbf{1} & 466 & 23.2 & 18 & 0.23  \\ & & & \textbf{2} & 632 & 8.6 & 5 & 0.11  \\

    \midrule
    \multirowcell{2}{\textbf{Experimental} \\ \textbf{Data}} & \multirowcell{2}{96} & \multirowcell{2}{-3.8} & \textbf{1} & 424 & 20.3 & 17 & 0.21  \\ & & & \textbf{2} & 653 & 9.8 & 5 & 0.10 \\

    \bottomrule
  \end{tabular}
  \caption{Comparison of electron properties retrieved from simulated (with ideal Gaussian and reconstructed laser profile) and experimental $\theta_x - E$ spectra shown in Fig.~\ref{fig:spectra_sim}. 
 $Q_{tot}$ is the total charge on each Fig.~\ref{fig:spectra_sim} spectrum, \textcolor{black}{$\theta_{x,max}$ the central divergence of the angular distribution},  while the "peak" labelled quantities refer to the $dQ/dE$ profiles that result from  selection of $dQ/dEd\theta_x$ within a $\pm 3$ mrad range around the maximum $dQ/d\theta_x$. Boldface denotes specific peak labels identified in Fig. 6 (a)-(c) $dQ/dE$ profiles. $\Delta E_{peak}/E_{peak}$ is the energy spread FWHM for each peak,  $Q_{peak}$ is the FWHM charge contained within each peak and $dQ/dE \,max.$ is the maximum $dQ/dE$ value reached within each peak.} 
  \label{fig:table_spectra_sim}
\end{table*}

These results show that a Gaussian driver Fig.~\ref{fig:spectra_sim}.(a) gives rise to a wide, high-energy, high-charge electron spectrum, peaked spatially on the laser axis. Using a realistic laser driver generates an off-axis electron beam with lower energy, lower charge, a \textcolor{black}{structure in agreement with experimental results as shown in Fig.~\ref{fig:spectra_sim}.(b) and (c) to (e), where all these spectra exhibit an off-axis $dQ/d\theta_x$ profile centered toward negative values.}
\textcolor{black}{The final spectrum contains 20 \% (0 \%) of self-injected electrons for the Gaussian case (realistic cases).}

Table~\ref{fig:table_spectra_sim} provides quantitative data for comparison of electron beam properties for the 3 cases shown in Fig.~\ref{fig:spectra_sim}.(a) to (c). The total charge $Q_{tot}$ is summed between $\pm 20$ mrad and $300 - 900$ MeV. We define the exit angle $\theta_{x,max}$ as the angle at which the maximum $dQ/d\theta_x$ is reached. Within $\pm 3$ mrad, peaks with center energy $E_{peak}$, FWHM width $\Delta E_{peak}/E_{peak}$, and FWHM charge $Q_{peak}$, are identified based on minimum peak prominence $\delta_{dQ/dE} = 0.05$ pC/MeV and minimum base to base width $\delta_E=45$ MeV.
Values of Table~\ref{fig:table_spectra_sim} show that the results of the simulation with reconstructed laser field profile are in good agreement with the detailed electron beam structure measured in experiment, while the Gaussian simulation results in electron beam characteristics significantly different from the measured ones. \textcolor{black}{Two major differences are $Q_{tot}$ value, which is $45\,\%$ higher than the experimental data for the Gaussian case, and the overall $dQ/dE$ structure which has a highest peak energy of $875$ MeV against $653$ MeV in the experimental data}. The charge difference stems from the quality of the laser angular intensity profile as it is the only input difference between Gaussian and Laguerre-Gauss simulations. 

The analysis of the evolution of  laser symmetry during the propagation coupled to electron injection is discussed in the next section. 

\subsection{Evolution of the laser asymmetry and effects on electron dynamics}
\begin{figure*}[ht!]
    \includegraphics[width=\textwidth]{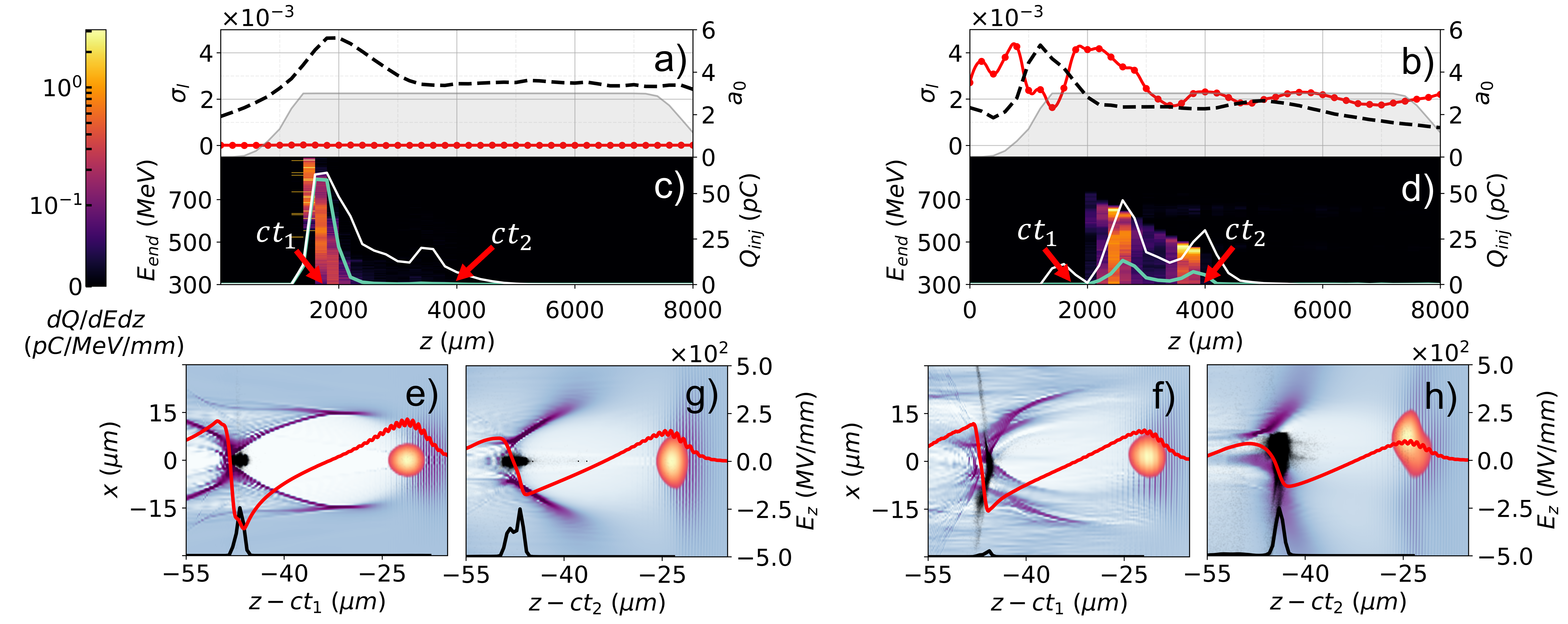}
    \caption{Simulated evolution of laser pulse and electron bunch characteristics for the ideal Gaussian laser profile (columns on the left) and the realistic laser field profile [columns on the right]. Panels (a) and (b): laser asymmetry function  $\sigma_l$ (red dots) and  spline approximated asymmetry function curve (red curve) as functions of position on the propagation axis;   peak normalized laser potential $a_0$  (black dashed-line), and normalized longitudinal unperturbed plasma density profile (grey area). Panels (c) and (d): injected charge $Q_{inj}$ (vertical axis on the right)  as function of position (all electrons plotted as a white line, and angularly selected $\pm 3$ mrad electrons as green line); The panels (c) and (d) also show the spectral charge density $dQ/dE_{end}dz$  of electrons (black-red-yellow colormap) with final energy $E_{end}$ for the angular selection as a function of the injection position $z$.
    Panels (e) and \textcolor{black}{(g)} [\textcolor{black}{(f)} and (h) for the simulation with realistic laser field]: transverse slice of the normalized charge density perturbation $\rho/en_0$ on the $z-x$ plane at $ct_1 = 1800$ $\mu$m and resp. $ct_2 = 4000$ $\mu$m; in these panels the electron macro-particles positions are shown as small black dots (with their size pondered by their individual charge). Superposed in these panels is the laser driver intensity's normalized envelope (orange colormap); in the same panels, the black and red lines correspond respectively to the density and the longitudinal electric field $E_z$ on the axis of maximum laser intensity.}
    \label{fig:physics_sim}
\end{figure*}

Figure~\ref{fig:physics_sim} shows the simulated evolution of the laser beam and electron bunch characteristics in the $xOz$ plane for the case with ideal Gaussian laser profile (left-hand column) and the case with the reconstructed laser field (right-hand column). It can be inferred from Fig.~\ref{fig:physics_sim} that the evolution of laser asymmetry and maximum field amplitude define the conditions for electron injection and the dynamics of electrons during the acceleration process in the plasma cavity. 

Figures~\ref{fig:physics_sim}.(a) and (b) show the evolution of the laser driver in the plasma density profile. As the realistic simulated transverse distribution is asymmetric with respect to the focal plane and changes before and after $z=z_f$ [Fig. \ref{fig:GSA_fit}], different spatial distortions occur during  non-linear self-focusing  inside the plasma. 

To quantify the deviation from cylindrical symmetry of the transverse fluence $F(r,\theta)$ and track its evolution throughout the simulated propagation, we define an asymmetry coefficient $\sigma_l$~\cite{Dickson2022} as :
\begin{equation}
    \sigma_l = \int_{r=0}^{r_{max}}\mathop{r}\mathop{dr}\sqrt{\int_{\theta = 0}^{2\pi} 
    \left( \mathop{d\theta}\left(f(r,\theta) - \overline{f}(r)\right) \right)^2 } \, ,
\end{equation}
where the normalized laser fluence $f(r,\theta)$ is defined as
\begin{equation}
    f(r,\theta)=\dfrac{F(r,\theta)}{\int_{r=0}^{r_{max}}\int_{\theta = 0}^{2\pi}\mathop{r} \mathop{dr}\mathop{d\theta}F(r,\theta) },
\end{equation} and $\overline{f}(r)$ is the mean normalized fluence over $\theta$. The integral origin $r=0$ is defined as the position of the fluence maximum. By definition,
$\sigma_l$ converges toward $0$ for a cylindrically symmetric fluence profile. 

This asymmetry coefficient $\sigma_l$ is plotted for the Gaussian laser profile and realistic laser profile in Fig.~\ref{fig:physics_sim}.(a) and (b) respectively as a function of position along propagation axis. Figure~\ref{fig:physics_sim}.(a) confirms that the Gaussian profile is angularly symmetric.
In this ideal case $\sigma_l$ undergoes variations between $0.5e{-5}$ and $2.5e{-5}$. In comparison, its variations in the simulation with realistic laser field profile are on a scale $\mathrm{10^2}$ times larger.
In the realistic case (Fig.~\ref{fig:physics_sim}.(b)), the amplitude of the symmetry coefficient drops by $67\,\%$ between $600$ and $2000$ $\mu$m, the focal plane in the plasma being in the middle of this area. This reduction of $\sigma_l$ is simultaneous with the increase in $a_0$, showing that  tight focusing of the laser reduces  its imperfections. In the realistic case, $a_0$ reaches values similar to the Gaussian $a_0$ through the first self-focusing due to a near Gaussian shape in its focal plane (Fig.~\ref{fig:GSA_fit}.(d)). However, the asymmetry of the intensity around the focal plane and relatively short typical variation length ($z_R = 1$ mm) are responsible for a $1$ mm shift of the maximum in comparison to the Gaussian symmetric pulse [Fig.~\ref{fig:physics_sim}.(a) and (b)].\\

Figures~\ref{fig:physics_sim}.(c) and (d) show line profiles of injected charge $Q_{inj}$ (right hand-side vertical axis), and  spectral charge density $dQ/dE_{end}dz$  of electrons (black-red-yellow histogram) with final energy $E_{end}$, as  functions of injection position $z$. In other words, this spectrum describes distribution of the initial $z$ positions of electrons for given final energies $E_{end}$. 

The simulation with ideal Gaussian laser profile [Fig.~\ref{fig:physics_sim}.(c)] 
shows a correlation between final energy and injection position as electrons trapped earlier gain more energy over the interaction distance.
The injection of electrons in $\pm 3$ mrad around axis (green line) occurs inside a $400$ $\mu$m window centered around the maximum laser amplitude $a_{max}$ at $ct_1=1800$ $\mu$m. The green injected charge in this region (Fig.~\ref{fig:physics_sim}.(c)) is maximal because the process occurs over $0.4z_R$ within the density plateau, in the portion where $a_0\simeq a_{max}$. Electrons injected at the start of this region see a longer accelerating length than the ones at the end, which results in decreasing final energy depending on the injection position. The resulting spectrum is evenly spread between $300$ and $900$ MeV as observed in Fig.~\ref{fig:spectra_sim}.(a).

For a realistic laser driver, the injected charge spectrum (Fig~\ref{fig:physics_sim}.(d)) exhibits multiple injection positions.
The higher energy electrons are injected $1$ mm after $ct_1$, when $a_0$ has dropped to $2.5$, and contribute to the high energy peak centered on $632$ MeV in Fig.~\ref{fig:spectra_sim}.(b). The lower energy peak at $466$ MeV in Fig.~\ref{fig:spectra_sim}.(b) is composed of electrons trapped at different positions: $1$ mm after $ct_1$ and at $ct_2$, which explains why this peak (Index 1 of Figs.~\ref{fig:spectra_sim}.(b) and (c)) has a relatively higher energy spread than peak 1 for Gaussian driver (Table ~\ref{fig:table_spectra_sim}).

The dynamics of electron  injection results from the evolution of the laser and plasma cavity. Snapshots of the first plasma cavity behind the laser pulse are shown for the Gaussian driver in Fig.~\ref{fig:physics_sim}.(e) and (g), and for a realistic driver in Fig.~\ref{fig:physics_sim}.(f) and (h),   at  $ct_1=1800$ $\mu$m and $ct_2 = 4000$ $\mu$m respectively. At $ct_1$, for the ideal Gaussian laser simulation, injection of electrons on axis has just begun in a perfectly symmetric bubble, while for  the realistic case most of the already injected electrons are greatly defocused and spread from $-30$ to $30$ $\mu$m. The accelerating fields are similar for Fig.~\ref{fig:physics_sim}.(e) and (f). However, in the realistic case,  electrons ionized early  are desynchronized with the trapping portion of the bubble due to off axis laser fluence fluctuations resulting in a $3$ $\mu$m shorter bubble compared to the Gaussian case. This prevents continuous injection and the generation of electrons with energy above $700$ MeV [Fig.~\ref{fig:spectra_sim}.(a) and (b)].

Comparison of the two cavity structures  at $ct_2$ (Fig.~\ref{fig:physics_sim} (g) and (h)), provides insight on the effects of transverse asymmetry evolution on the acceleration process.

In the ideal Gaussian laser case, most of the initially injected charge is accelerated, and the injection process remains continuous.

The transverse centroid variations and stronger defocusing with the realistic laser profile induce important losses throughout the interaction process, which enables  trapping of up to $10$ pC  around $ct_2$, in an off-axis bubble with half of the accelerating maximum compared to a Gaussian cylindrically symmetric driver.

\begin{figure}[ht!]
    \includegraphics[width=\linewidth]{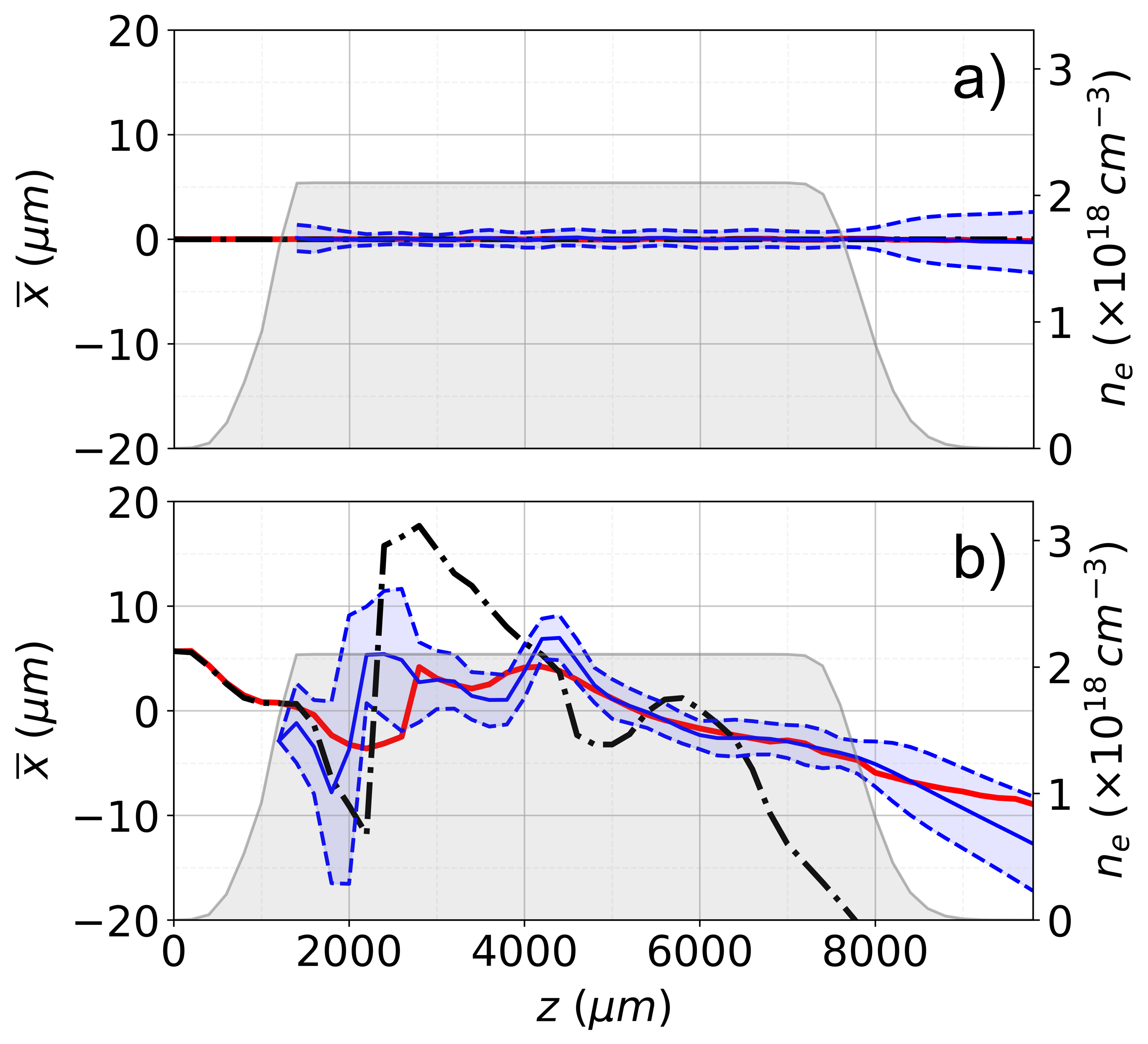}
    \caption{ Laser beam  (solid red line)  and electron bunch  (solid blue line) positions in the xOz plane for  (a)  Gaussian case and (b) Laguerre-Gauss case.
    The blue dashed lines
    are the $ \pm$ RMS $\sigma_x$ size of the selected bunch.
    The laser position is defined as the centroid of the fluence distribution in an area of radius $r_{eff}$ around the fluence maximum. For reference,  the dash-dotted black line represents the position of the laser beam in vacuum and the  grey area is the longitudinal density profile (right-end side vertical axis scale).}
    \label{fig:symmetry_sim}
\end{figure}

Laser and electron beam positions in the $xOz$ transverse plane are plotted in Fig.~\ref{fig:symmetry_sim} as functions of the position along the propagation axis in vacuum. The ideal Gaussian laser trajectory is centered on the laser axis in vacuum, closely followed by the electron beam trajectory. The electron beam transverse size remains relatively constant in the plasma and grows symmetrically around $ x=0$ after plasma exit.
For a realistic laser driver, the laser trajectory [Fig.~\ref{fig:symmetry_sim}.(b) ]in vacuum oscillates around the $ x=0$  axis. Self-focusing in the plasma  lowers the amplitude of the displacement off-axis (compare red line and black dashed line). 

After electron trapping, $ z > 2000$ $\mu$m, the electron bunch trajectory  is centered on the laser centroid, and its standard deviation reaches up to $10$ $\mu$m. This behavior clearly demonstrates the impact  of the asymmetry of the transverse laser driver on the pointing of the electron beam at the exit of the plasma.

\subsection{Charge fluctuations due to laser intensity distribution}

Simulations were performed to analyze the effects of variations of the intensity profile on the resulting charge. The  laser electric field was reconstructed for  3 reference profiles with the modified GSA algorithm described in Appendix \ref{appendix:GSA_fit} and the remaining simulation input parameters are the same as in Fig.~\ref{fig:spectra_sim}.(b) (reported in section \ref{sec:realistic_sim}). \textcolor{black}{The density was varied between $1.7\times\mathrm{10^{18}}$ and
$2.3\times\mathrm{10^{18}}$ cm$^{-3}$}.

To characterize the laser energy distribution for each profile, we define the following effective energy ratio :
\begin{equation}
    E_{ratio} = \dfrac{1}{N_z}\sum_{z=-z_R}^{+z_R}\dfrac{\int\displaylimits_{r=0}^{r_{eff}}\int\displaylimits_{\theta=0}^{2\pi}\mathop{r}\mathop{dr}\mathop{d\theta} F(r,\theta, z)}{\int\displaylimits_{r=0}^{r_{max}}\int\displaylimits_{\theta=0}^{2\pi}\mathop{r}\mathop{dr}\mathop{d\theta} F(r,\theta, z)} \, ,
    \label{eq:Eratio}
\end{equation}
with $z$ being the propagation position of the laser with respect to $z_f$, $N_z$ the number of evenly spaced positions $z$ used to perform the summation in the Rayleigh range $[-z_R, ..., 0, ..., z_R]$ and $F(r,\theta, z)$ the fluence at each propagation position in vacuum. The integral origin is defined as the position of the fluence maximum. This ratio quantifies the average effective energy in the characteristic divergence boundaries $[-z_R,\,z_R]$ of the laser propagation axis, and within an area of radius $r_{eff}$. Improving this ratio increases the portion of energy usable for the injection of electrons.

\begin{figure}[ht!]
    \includegraphics[width=0.9\linewidth]{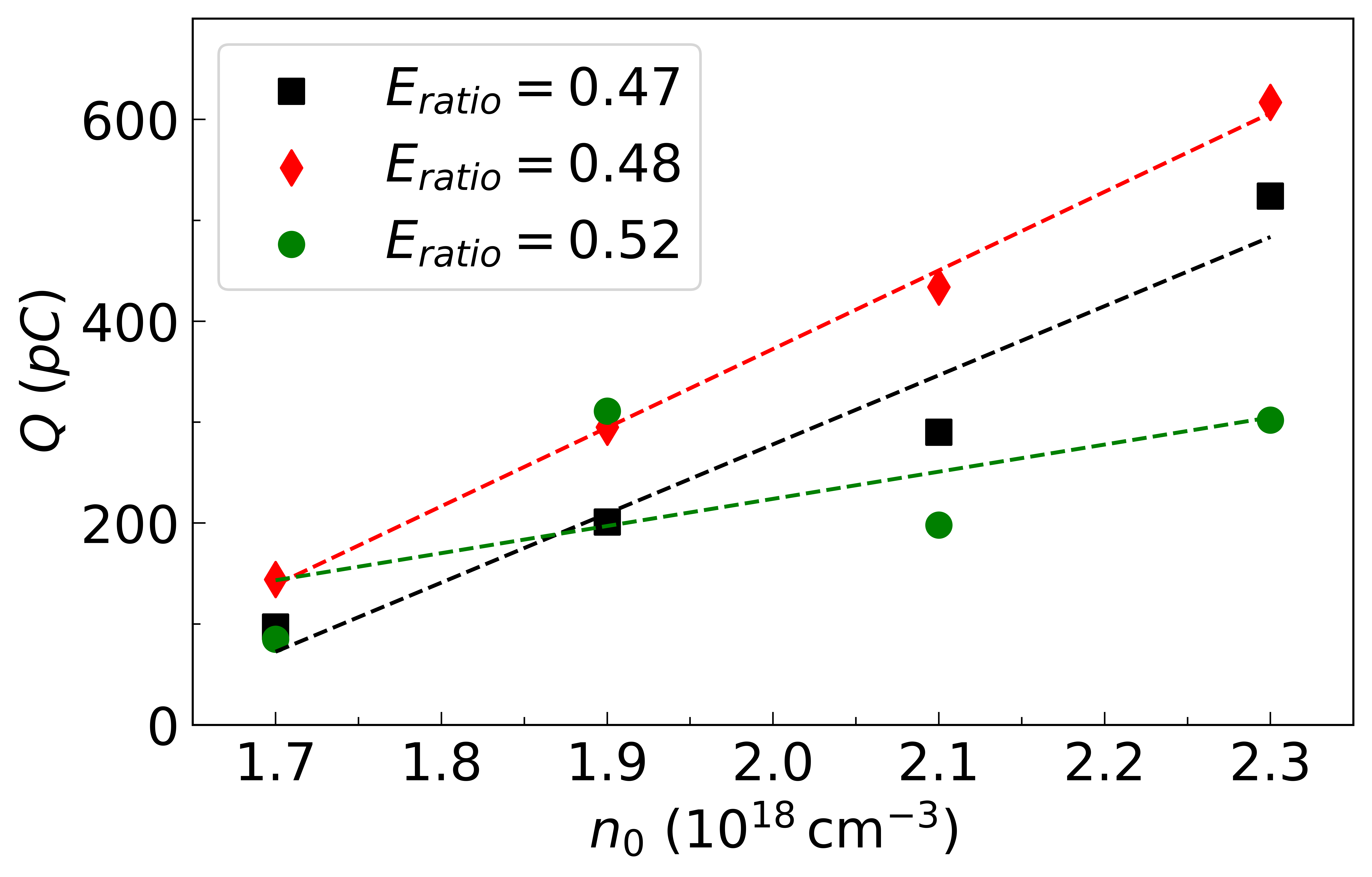}
    \caption{Total injected bunch charge $Q$ for 3 different simulated laser profiles as a function of $n_0$. For each profile, the corresponding $E_{ratio}$ is calculated by eq. (\ref{eq:Eratio}), setting $N_z=21$:
    black squares profile with $E_{ratio}=0.47$ - red diamonds profile with $E_{ratio}=0.48$ - green circles profile with $E_{ratio}=0.52$. For each profile, the dashed curve is the linear regression of the data points.}
    \label{fig:Q_En0}
\end{figure}

The total accelerated charge $Q$ is calculated within the whole $E-\theta_x$ space and plotted in Fig. \ref{fig:Q_En0} as a function of $n_0$ for the 3 laser profiles and  densities ranging from $1.7$ to $2.3\times10^{18}$ cm$^{-3}$.
In this regime, within the chosen set of parameters, the total charge $Q$ remains lower than the theoretical value calculated for a matched laser  for most of the input densities $n_0$. This is a consequence of the non-Gaussian laser transverse distributions, and of the fact that the laser spot size is not matched with the bubble radius. Comparing the 3 values simulated for $n_0=2.1\times10^{18}$ cm$^{-3}$, while $E_{ratio}$ fluctuates between $0.47$ and $0.52$, the total charge $Q$ varies from  $198$ to $434$ pC. A relative effective energy fluctuation $\delta E_{ratio}/E_{ratio} = 5\,\%$ leads to a relative charge fluctuation $\delta Q / Q = 38\,\%$. For the laser profile resulting in the highest total charge, i.e. $E_{ratio}=0.48$, increasing the density $n_0$ from $1.9\times10^{18}$ to $2.1\times10^{18}$ cm$^{-3}$ increases $Q$ from $295$ to $434$ pC. A relative density fluctuation of  $5\,\%$ leads to a relative charge fluctuation $\delta Q / Q = 19\,\%$.
This shows that taking into account small errors either on the gas density value or on the calculated effective energy ratio leads to significant charge fluctuations, of  the same order of magnitude  as the one measured experimentally (Fig.~\ref{fig:spectra_exp}).

The averaged calculated charge measured within the same  boundaries as experimental data is $113$ pC ($\overline{Q}_{tot}=95$ pC in experiment), with an RMS relative charge fluctuation of 56$\%$ ($\delta \overline{Q}_{tot} / \overline{Q}_{tot} = 47\,\%$ in experiment). 
There is a good agreement between the average simulated and experimental charge for the same central density $\overline{n}_0=2.2\times10^{18}$ cm$^{-3}$. 
In conclusion, this analysis shows that improving the quality and the stability of the laser energy distribution and the stability of the plasma density are crucial to achieve stable and high-energy, high charge electron spectra. 

\section{Conclusions}
Experiments were carried out during  the commissioning phase in the long focal area of the APOLLON laser facility, to study the influence of the laser beam properties on the quality of electron beams generated by ionization injection and laser wakefield acceleration in a gas cell. A detailed analysis of the laser beam in the transverse plane was performed using fluence measurements in the focal volume. Particle in Cell simulations were performed with a reconstruction of the laser field obtained from the measured laser fluence map as input, leading to electron spectra in close agreement with experimental results, for the accelerated charge, energy distribution and pointing of the electron beam at the exit of the plasma. The presented results also show that this degree of quantitative agreement can be found without using computationally demanding full 3D simulations.

These high fidelity simulations rely on the calculation of the \textcolor{black}{laser} electric field from experimental data providing  the fluence at different positions along the propagation in vacuum.

An iterative method based on a modified version of the Gerchberg-Saxton algorithm~\cite{gerchberg1972practical}, which allows the reconstruction of a realistic \textcolor{black}{laser} electric field based on a collection of fluence images in vacuum (Fig.~\ref{fig:GSA_fit}), was used. 
The implementation of asymmetric laser drivers leads to better agreement of the simulation output bunch characteristics to measured experimental data in comparison to simulations using a perfect Gaussian driver (Fig.~\ref{fig:spectra_sim} and Table~\ref{fig:table_spectra_sim}). 

These realistic simulations highlight the effects of laser field spatial characteristics (centroid fluctuations and asymmetry quantified by $\sigma_l$) on the injection and acceleration of the electrons (Fig.~\ref{fig:physics_sim} and Fig.~\ref{fig:symmetry_sim}). 
The symmetry degradation from a Gaussian laser driver leads to a loss of both peak energy and total accelerated bunch charge (Table~\ref{fig:table_spectra_sim}), which could be mitigated through optimization of the gas cell density characteristics.\\
The impact of shot-to-shot fluctuations of the laser transverse distribution and of plasma density on the accelerated bunch charge, have been quantified and the source of charge fluctuations in experiments identified. The stabilization of these fluctuations would lead to an improved stability of the produced electron spectra.

\section*{Acknowledgments}
 We acknowledge the resources and assistance of the computing center MesoLUM managed by ISMO (UMR8214) and LPGP (UMR8578), University Paris-Saclay (France).  Experimental results were achieved using APOLLON Research Infrastructure and we gratefully acknowledge the work of the Laboratoire pour l'Utilisation des Lasers Intenses (UMR7605) staff during planning and execution of the experiment. APOLLON facility was  partially funded by Equipex Cilex (Centre interdisciplinaire lumière extrême) grant N$^\circ$ ANR-10-EQPX-25-01, and by region Ile-de-France.



\appendix

\section{Modified Gerchberg-Saxton algorithm (GSA) to model the laser driver}\label{appendix:GSA_fit}

In the experimental campaign described in this work, only the fluence maps $F(x,y,z)$ of the laser pulse at specific distances $z_k$, with $k=0,1,...,k_{\mathrm{max}}$ from the focal spot were measured.

A Gaussian temporal profile was assumed for the laser pulse, with the measured FWHM duration in intensity $\tau_{FWHM}$. Under this hypothesis, the linear relation between the peak fluence $F_0$ and peak intensity $I_0$ is :
\begin{equation}
    F_0 = \dfrac{\tau_{FWHM}}{2}\sqrt{\dfrac{\pi}{\log 2}} I_0.
\end{equation}

Given this linear relation between fluence $F(x,y,z)$ and intensity $I(x,y,z)$, an electric field at position $z$ can be defined from a phase map $\psi(x,y)$ and a fluence $F(x,y,z)$:
\begin{eqnarray}\label{eq:e_reconstruction}
\mathrm{E}(x,y,z) = \mathrm{E}[F(x,y,z),\psi(x,y)] = \nonumber \\ =\sqrt{I(x,y,z_0)}\exp{i\psi(x,y)}.
\end{eqnarray}

To initialize the realistic PIC simulations of this work, a reconstruction of the laser electric field at a given plane was necessary. Since only the fluence (and thus intensity) maps at multiple planes were known experimentally, to reconstruct the \textcolor{black}{laser} electric field using Eq. \ref{eq:e_reconstruction} the field phase map had to be reconstructed. 

For this purpose, a modified implementation of the Gerchberg-Saxton algorithm (GSA) \cite{gerchberg1972practical} was used to find the field phase $\psi(x,y)$ at $z_0$ and thus the \textcolor{black}{laser} electric field $\mathrm{E}_{GSA,z_0}(x,y,z_0)=\mathrm{E}[F(x,y,z_0),\psi(x,y)]$, from the available data on the the fluence $F(x,y,z_0)$ at $z_0$ and other planes, each referred to as $z_k$.

In the following  a simplified description of the field reconstruction algorithm used in this work  is reported.

This version of the GSA aims at finding a reconstruction of the laser field at $z_0$ through an expansion in Hermite-Gauss (HG) modes \cite{pampaloni2004Gaussian}, using the measured $F(x,y,z_k)$ fluence maps. Since the propagation of the HG modes $\mathrm{HG_{n,p}}$ at $z_k$ is analytically known, this reconstruction is equivalent to finding the estimated HG expansion coefficients $D_{n,p}$ (and thus the corresponding field phase map). The indices $n$,$p$ denote the order of the HG mode along the $x$, $y$ axis respectively.
 
At each iteration $iter$ of the algorithm, the estimated expansion in HG modes is propagated from the position $z_0$ to $z_{k_{\mathrm{max}}}$, with an improvement of the estimate at each of the intermediate positions $z_k$. This update of the estimated coefficients for the HG expansion uses the estimated coefficients from the previous measurement plane at $z_{k-1}$ and the measured fluences $F(x,y,z_k)$. As in the original GSA formulation \cite{gerchberg1972practical}, the estimated propagated phase is combined with the fluence at the measurement planes in the calculations.

After one iteration ends, the procedure is repeated starting from $z_0$, using the coefficients (and thus the phase map) estimated from the previous iteration.

The implementation of the modified GSA used for this work can be described by the following pseudocode:
\begin{itemize}
\item Find an initial estimate of the coefficients $D_{n,p}$ projecting the intensity corresponding to the fluence $F(x,y,z_0)$ on the $\mathrm{HG_{n,p}}(x,y,z_0)$ modes at $z_0$. In the following we denote the projection of a function $f(x,y,z)$ on the HG modes with the notation $\mathrm{Proj}$:
\begin{equation}
D_{n,p}=\mathrm{Proj}[f(x,y,z),\mathrm{HG_{n,p}}(x,y,z)].\\
\end{equation}
\item For $iter=0 \rightarrow N_{\mathrm{iter}}$ and for $k=0 \rightarrow k_{\mathrm{max}}$:
\begin{itemize}
\item define the propagated field as 
\begin{equation}
\mathrm{E}_{GSA,z_k} (x,y,z_k)=\sum_{n,p} D_{n,p}\cdot \mathrm{HG_{n,p}}(x,y,z_k);
\end{equation}
\item find the phase map $\psi$ as:
\begin{equation}
\psi(x,y)=\arg{(\mathrm{E}_{GSA,z_k} (x,y,z_k))};
\end{equation}
\item combine the measured fluence $F(x,y,z_k)$ with the phase $\psi(x,y)$ to find the function $\mathrm{E'}_{GSA,z_k}$ using Eq. \ref{eq:e_reconstruction}:
\begin{equation}
\mathrm{E'}_{GSA,z_k} (x,y,z_k)= \mathrm{E}[F(x,y,z_k),\psi(x,y)];
\end{equation}
\item combine the previous estimate of the HG coefficients with those obtained from the projection of $\mathrm{E'}_{GSA,z_k}$ on the HG modes at $z_k$:
\begin{eqnarray}
D_{n,p}=(1 - \alpha)\cdot D_{n,p}+\nonumber\\ 
+\alpha\cdot\mathrm{Proj}[\mathrm{E'}_{GSA,z_k},\mathrm{HG_{n,p}}(x,y,z_k)];
\end{eqnarray}

\end{itemize}

\end{itemize}

For this work the number of iterations was chosen as $N_{\mathrm{iter}}=10$. The coefficient $\alpha$ for the weighted sum of the previous and new HG expansion coefficients are chosen in order to obtain convergence.

Once the algorithm has performed the iterations passing through the measurements planes, using Eq. \ref{eq:e_reconstruction} the estimated field phase $\psi(x,y)$ can be easily found and combined with the measured fluence $F(x,y,z_0)$ of the laser to reconstruct its field $\mathrm{E}_{GSA,z_0} (x,y,z_0)=\mathrm{E}[F(x,y,z_0),\psi(x,y)]$.

In this algorithm, the mentioned projection of a function $f(x,y,z_k)$ on the HG modes at $z_k$ was defined as:
\begin{eqnarray}
\mathrm{Proj}[f(x,y,z_k),\mathrm{HG_{n,p}}(x,y,z_k)] =\nonumber \\
=\int_{-\Delta X/2}^{\Delta X/2}\int_{-\Delta Y/2}^{\Delta Y/2} \mathop{dx} \mathop{dy} ~ f(x,y,z_k)\times \nonumber \\
\times \mathrm{HG}_{n,p}^*(x, y, z_k),
\end{eqnarray}
where $\left( \Delta X,\Delta Y \right)$ are the data rectangular grid length along each axis.
The HG modes at $z_k$ are defined also using the origins $\left(x_{0,k},y_{0,k}\right)$ and the projection waists $w_{0x,y}$.
The latter are set sufficiently small to make the Hermite-Gauss modes decay within the projection integral boundaries, and sufficiently large to fit the part of the transverse intensity map further away from the origin. The values of $\left(x_{0,k},y_{0,k}\right)$ are optimized to improve the quality of the laser field reconstruction, as will be described in a future work.

The field reconstruction obtained with described phase retrieval algorithm would be sufficient to initialize the laser pulse in a realistic PIC simulation in 3D Cartesian geometry. However, for the simulations in quasi-3D geometry \cite{LIFSCHITZ20091803} of this work, which use a cylindrical grid, a further decomposition in Laguerre-Gauss modes is necessary, as described in Section \ref{sec:pic_sim_setup}.

\bibliography{apssamp}
\end{document}